\documentclass[journal]{IEEEtran}
\pdfoutput=1
\ifCLASSINFOpdf
   \usepackage[pdftex]{graphicx}
\else
\fi
\begin{document}
%
\title{Performance of bare high-purity germanium detectors in liquid argon for the GERDA experiment}
%
%
%

\author{M. Barnab\'e Heider, C. Cattadori, O. Chkvorets, A. Di Vacri, K. Gusev, S. Sch\"onert, M. Shirchenko
\thanks{Manuscript received December 10, 2008. This work was supported by the Transregio Sonderforschungsbereich SFB/TR27 'Neutrinos and Beyond' by the Deutsche Forschungsgemeinschaft and by the INTAS grant 1000008-7996 of the European Commission.}        
\thanks{M. Barnab\'e Heider, O. Chkvorets and S. Sch\"onert are with the Max-Planck-Institut f\"ur Kernphysik, Saupferchckweg 1, D-69117 Heidelberg, Germany (corresponding author: Marik Barnab\'e Heider, Tel. +49 6221 516 812, Fax +49 6221 516 872, E-mail: heider@mpi-hd.mpg.de).}
\thanks{C. Cattadori and A. Di Vacri are with the Laboratori Nazionali del Gran Sasso, S.S. 17 bis km.18.910, 67010 Assergi (AQ), Italy.}
\thanks{O. Chkvorets is now with the Department of Physics, Laurentian University, Ramsey Lake Road, P3E 2C6 Sudbury, Ontario, Canada.}
\thanks{K. Gusev and M. Shirchenko are with the Russian Research Center Kurchatov Institute, 123182 Moscow, Russia and the Joint Institute for Nuclear Research, 141980 Dubna, Russia.}}

\maketitle

\begin{abstract}
The GERmanium Detector Array, GERDA, will search for neutrinoless double beta decay in $^{76}$Ge at the National Gran Sasso Laboratory of the INFN. Bare high-purity germanium detectors enriched in $^{76}$Ge will be submerged in liquid argon serving simultaneously as a shield against external radioactivity and as a cooling medium. In GERDA Phase-I, reprocessed enriched-Ge detectors, which were previously operated by the Heidelberg-Moscow and IGEX collaborations, will be redeployed. Before operating the enriched detectors, tests are performed with non-enriched bare HPGe detectors in the GERDA underground Detector Laboratory to test the Phase-I detector assembly, the detector handling protocols, the refurbishment technology and to study the long-term stability in liquid argon. The leakage currents in liquid argon and liquid nitrogen have been extensively studied under varying gamma irradiation conditions. In total three non-enriched high-purity p-type prototype germanium detectors have been operated successfully. The detector performance is stable over the long-term measurements. For the first time, performance of bare high-purity germanium detectors in liquid argon is reported.
\end{abstract}


\IEEEpeerreviewmaketitle

\section{Introduction}

\IEEEPARstart{T}{he} GERmanium Detector Array, GERDA, will search for neutrinoless double beta decay (0$\nu \beta \beta$) in $^{76}$Ge \cite{gerda} with the goal to investigate the nature of neutrinos and their mass. GERDA is located 1400 m (3800 m w.e.) underground at the Laboratori Nazionali del Gran Sasso (LNGS), Italy. To achieve extremely low background, the basic layout of GERDA follows ideas proposed several years ago in \cite{heusser}. Bare high purity germanium (HPGe) detectors enriched in $^{76}$Ge will be submerged in liquid argon. The high-purity liquid argon acts simultaneously as a cooling medium and as a shield against external radiation. By operating bare HPGe detectors, GERDA aims at an extremely low background and at an excellent energy resolution.

The experiment is foreseen to proceed in two phases. The first phase of GERDA will operate reprocessed enriched diodes from the past Heidelberg-Moscow (HdM) \cite{hdm} and IGEX \cite{igex} experiments. In total, 8 HPGe detectors (total mass $\sim$18 kg) enriched in $^{76}$Ge at 86 \% and 6 reprocessed natural HPGe detectors from the Genius Test-Facility \cite{genius} (total mass $\sim$15 kg) will be redeployed. The aim is to surpass the state-of-art in 0$\nu \beta \beta$ sensitivity during only a year of data-taking. This will be achieved through a 10-fold improvement in passive background reduction by shielding external radioactivity and minimizing the amount of material in detector support structure. A total background index of less than $10^{-2}$ cts/(keV$\cdot$kg$\cdot$y) in the Q$_{\beta\beta}$ range (2039 keV) should be achieved. Assuming an exposure of $\sim$15 kg$\cdot$y and an energy resolution of 3.6 keV, the expected number of background events is $<$0.5 counts. If no event is observed, a T$_{1/2}>3.0\cdot 10^{25}$ y (90\% C.L.) can be established with a detection efficiency of  95\%. This results in an upper limit on the effective neutrino mass of m$_{ee}<(0.3-0.9)$ eV, depending on the nuclear matrix elements used. In GERDA Phase II, new enriched diodes will be added to achieve 100 kg$\cdot$y of exposure within three years and the use of active background suppression techniques will be required to reduce the background index by one order of magnitude bellow 10$^{-3}$ cts/(keV$\cdot$kg$\cdot$y). Depending on the achieved physics results, an experiment with one ton of target mass and a background index of $<$10$^{-4}$ cts/(keV$\cdot$kg$\cdot$y) is considered in the framework of a new worldwide collaboration.

In preparation for GERDA, the Phase-I detectors are reprocessed by Canberra Semiconductor NV, Olen \cite{canberra}. Before submerging the enriched diodes in cryogenic liquid, the performance of bare HPGe detectors in liquid argon (LAr) and in liquid nitrogen (LN$_2$) has been investigated with non-enriched prototype detectors, which use the same technology as the Phase-I detectors. This work presents the results of an extensive experimental study with p-type HPGe detectors mounted in low-mass supports and operated in LAr and in LN$_2$ at the GERDA underground Detector Laboratory (GDL) at LNGS. The Phase-I detector assembly, the Phase-I detector technology and the GDL test facility are presented. The performance of the bare detectors is characterized with the study of the leakage currents (LC) in LAr and in LN$_2$ under varying $\gamma$ irradiation conditions and with the long-term stability measurements. Finally, the operations and measurements performed with the Phase-I enriched detectors are summarized.

\subsection{Experimental setup}
\begin{figure}[h]
		\begin{centering}
		\includegraphics[scale=0.08]{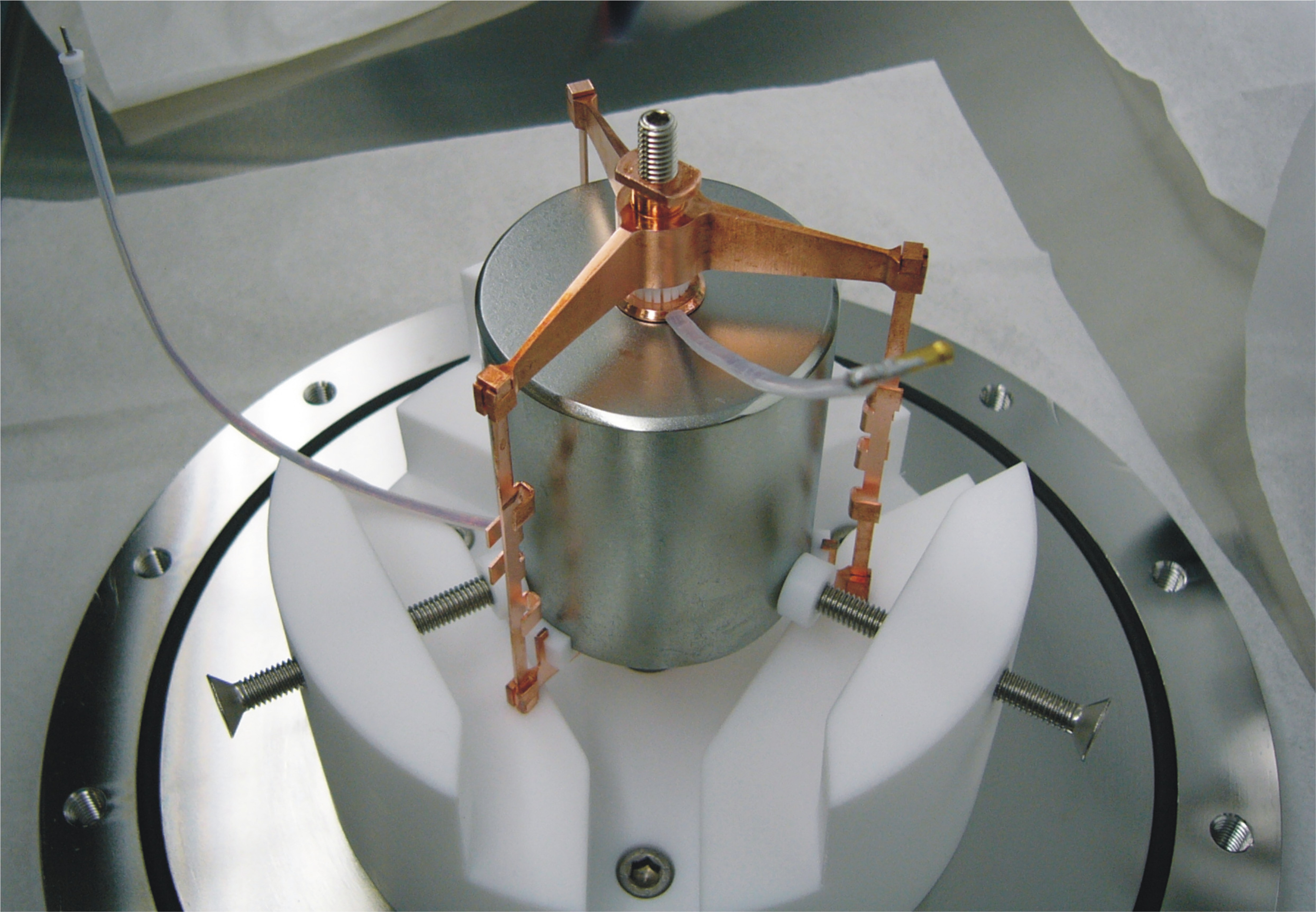}
	\caption{A prototype detector mounted in its low-mass holder.}
	\label{fig:protomounted}
	\end{centering}
\end{figure}

The HPGe detectors are mounted in a low-mass holder ($\sim$80 g), made of selected high radiopurity materials (low-activity copper, PTFE and silicon). Figure \ref{fig:protomounted} shows a prototype p-type detector mounted in its low-mass holder.
The detector is mounted with the bore hole side on the bottom and the high voltage (HV) contact side on top of the assembly. The detector assembly has been first tested in LN$_2$ at the detector manufacturer site. The mounting procedure, the signal and central HV contact quality, the mechanical stability and the spectroscopy performance of the Phase-I detector assembly have been tested successfully. The same energy resolution as obtained in a standard vacuum test cryostat (2.2 keV (FWHM) at 1.332 MeV) has been measured in LN$_2$ using a signal cable of $\sim$20 cm which connects the detector assembly with the warm front end electronics.

The Phase-I detectors were reprocessed by Canberra Semiconductor NV, Olen \cite{canberra} according to their standard technology. The detectors have a 'wrap around' n$^+$ conductive lithium layer which is separated from the signal contact by a groove covered by a passivation layer. Three prototype detectors using different passivation procedures of the groove have been tested. The first prototype (1.6 kg) had a full passivation, which covers the groove but also extends to the inner and outer surfaces on the bore hole side. The second prototype (2.5 kg) had a passivation layer limited to the groove only and the third prototype (2.5 kg) had no passivation layer.  

\begin{figure}[h]
		\begin{centering}
		\includegraphics[scale=0.06]{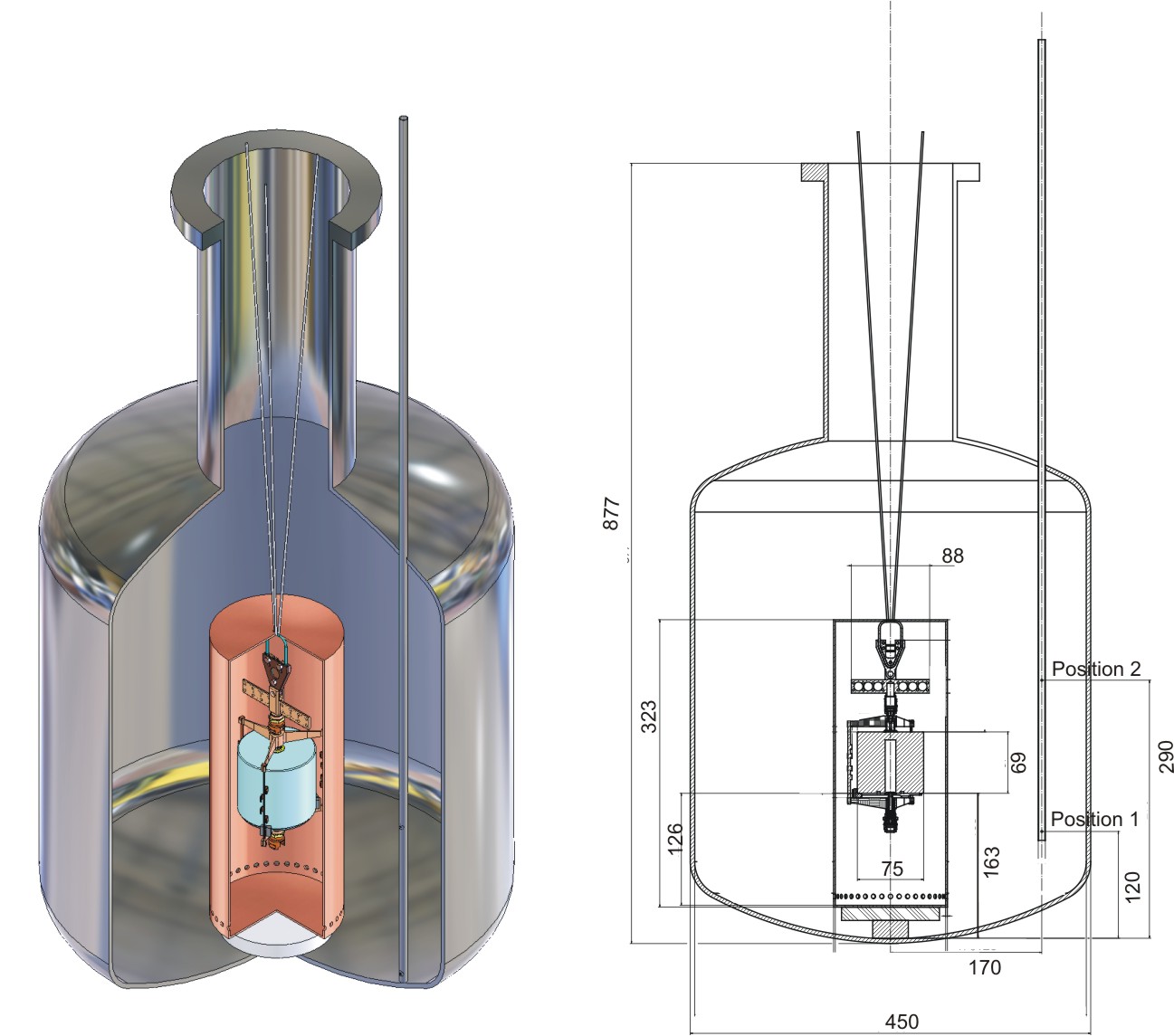}
	\caption{View and drawing of a LAr/LN$_2$ dewar in which the bare HPGe detectors are operated in the Gerda Detector Laboratory (GDL).}
	\label{fig:dewar}
	\end{centering}
\end{figure}

The bare detectors were operated in GDL, which is a clean room of level 10 000 equipped with a clean bench and a radon-reduced bench of level 10. The radon concentration of the laboratory ($\sim$10 Bq/m$^3$) is monitored and the humidity of the laboratory is kept low (30\%). The mounting of the detector assembly is performed inside the radon-reduced bench which is connected to a LAr/LN$_2$ detector test bench. The handling is carried out in a closed environment under N$_2$ atmosphere. Figure \ref{fig:dewar} shows a view and a drawing of the inner vessel of the 70 liter dewar in which the bare detectors are operated.
A copper cylinder is mounted inside the dewar to shield the detectors from the infrared light. The infrared radiation comes mainly from the neck of the dewar. In the present setup, the detector assembly is close enough to the dewar neck so it is sensitive to the infrared light. On the contrary, the detector array in GERDA will be sufficiently far away from the neck so no infrared shield will be needed. In order to have the infrared shield always submerged, the liquid argon is topped-up every week. The detector assembly is attached to the top of the infrared shield which is attached to the dewar flange. There are two LAr/LN$_2$ detector test facilities operational in GDL. The energy resolution obtained with both setups is 2.5 keV FWHM at 1.332 MeV, with a signal cable of $\sim$60 cm connecting the detector to the front end electronics. 

In Figure \ref{fig:dewar}, a tube for inserting a radioactive source in the proximity of the detector is shown.  Two source positions are indicated. At these positions, the distances between the source and the detector assembly are similar but the source in Position 1 irradiates mainly the LAr volume facing the bottom side (bore hole side) and the source in Position 2 irradiates mainly the LAr volume facing the top side (HV contact side) of the detector assembly. For the $\gamma$ irradiation measurements, a $^{60}$Co source with an activity of 44 kBq mounted on a steel wire was placed $\sim$20 cm away from the middle of the detector.

\subsection{Measurements}
\subsubsection{Detector handling summary}
During the first year, the LAr/LN$_2$ test bench and the detector assembly were optimized for cleanliness, optimal mounting procedure, low LC measurement and spectroscopy performance. Some modifications were done to the detector assembly and to the detector test stand to improve the detector performance. While the diode is mounted in its low-mass support and cooled down in the test stand dewar, measurements are performed to monitor the conditions of the detector assembly. The quality of the signal and HV contacts is measured before, during and after the cooling process. Subsequently, the LC and the energy resolution are recorded as the HV is applied to the detector. 

Within the first year of testing, $\sim$50 cooling-warming cycles have been performed with the first prototype to do mounting and/or electronics modifications. The warming-up procedure is very quick and none of our detectors has ever been damaged by this procedure. However, it has been observed that improper handling can damage the passivation layer of the detector. It leads to an intolerable increase of the LC which can be healed by a repair of the passivation layer.

\subsubsection{Leakage current (LC) increase in response to $\gamma$-radiation}
\begin{figure}[h]
		\begin{centering}
		\includegraphics[scale=0.25]{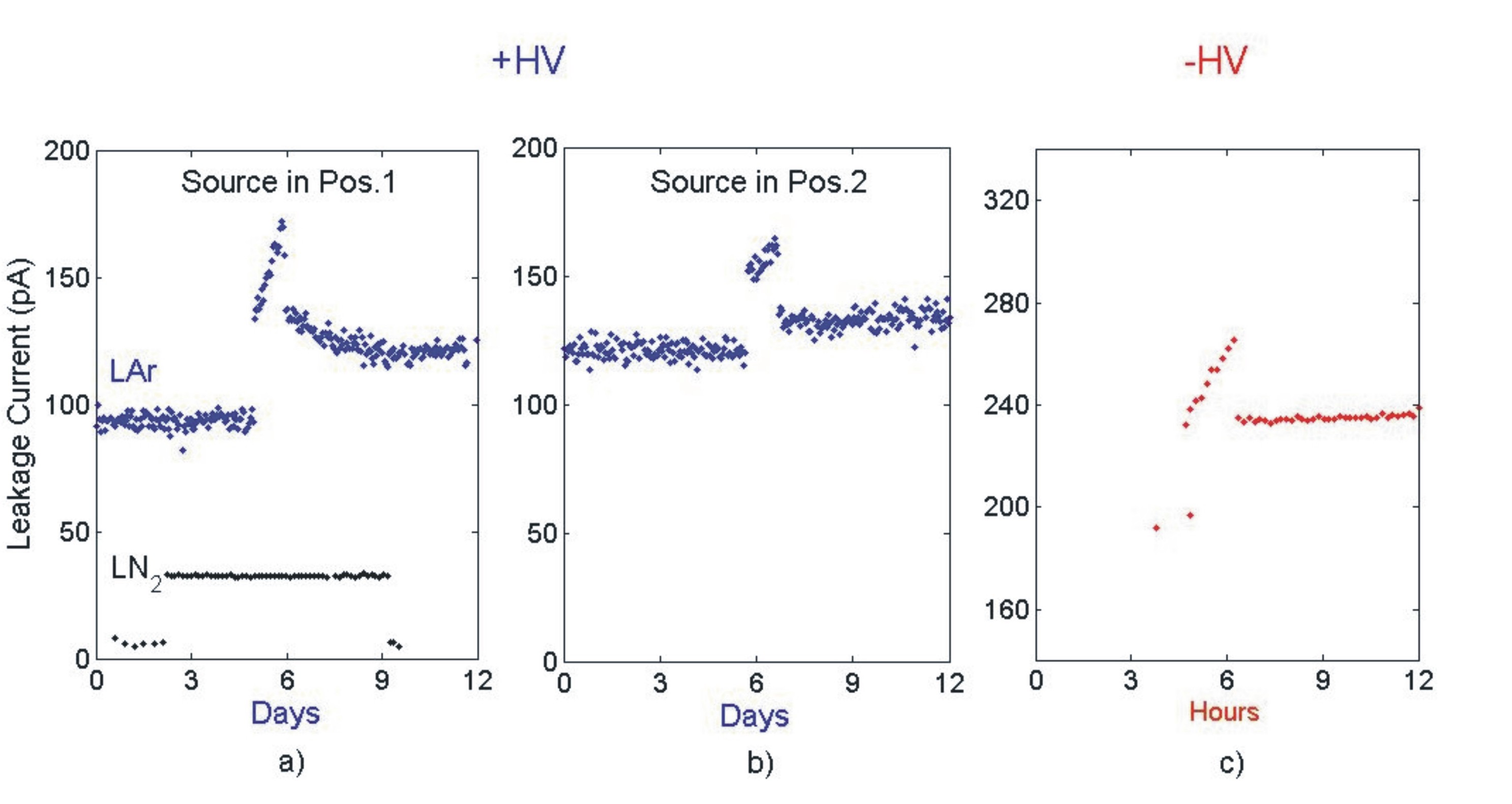}
	\caption{a) Gamma-radiation induced LC of the first prototype in LAr. The bulk current step ($\sim$35 pA) as the source is inserted in the setup is followed by a continuous increase of the LC. After one day of irradiation, the source is removed and the LC stabilizes at a higher value than prior to the irradiation ($\Delta$LC$\sim$30 pA). No increase of the LC is observed in LN$_2$ after one week of irradiation. b) Gamma-radiation induced LC with the source in Pos.2. c) Gamma-radiation induced LC with - HV applied to the detector.}
	\label{fig:increaseLC}
	\end{centering}
\end{figure}

Figure \ref{fig:increaseLC} a) presents the increase of the LC observed with the first prototype operated in LAr and exposed to $\gamma$-radiation. The LC is monitored every minute and the data are averaged over one hour periods. The expected spontaneous increase/decrease of the bulk current (I$_{Bulk}$) as the source is inserted/removed, (I$_{Bulk}=\frac{C<E>}{2.95 eV/e^-}$ where C is the counting rate (Hz), $<E>$ is the average energy deposited in the detector and 2.95 eV is the energy necessary to produce an e$^-$-hole pair in germanium at 80 K) is clearly seen.
The step is followed by an unexpected continuous increase of the LC. When the source is removed, the LC decreases and stabilizes at a higher value than prior to the irradiation. The LC increasing rate is not linear in time, but shows rather an exponential-like behaviour. Gamma irradiation of the same detector assembly in LN$_2$ showed no increase of the LC.

Irradiations with the $\gamma$ source in Position 1 and in Position 2 (Figure \ref{fig:dewar}) show that for the same distance source-detector, the LC increase is stronger if the source is located at the bottom of the detector assembly (Position 1) irradiating mainly the passivation layer side.

To investigate the mechanism behind the $\gamma$-radiation induced LC, measurements were performed with inverse HV polarity: - HV is applied to the p+ contact and the n+ contact is grounded. Inverting the polarity does not change the electric field inside the detector but it does change the field in the surrounding of the detector assembly. It has been observed that the LC is higher and the $\gamma$-radiation induced increase of the LC is faster with - HV (Figure \ref{fig:increaseLC} c)). The increase of the LC per day of irradiation with + HV is similar to the increase per hour of irradiation with - HV. In summary, the LC increase rate depends on the distance between the source and the passivated groove and on the electric field in the surrounding LAr volume.

\begin{figure}[h]
		\begin{centering}
		\includegraphics[scale=0.2]{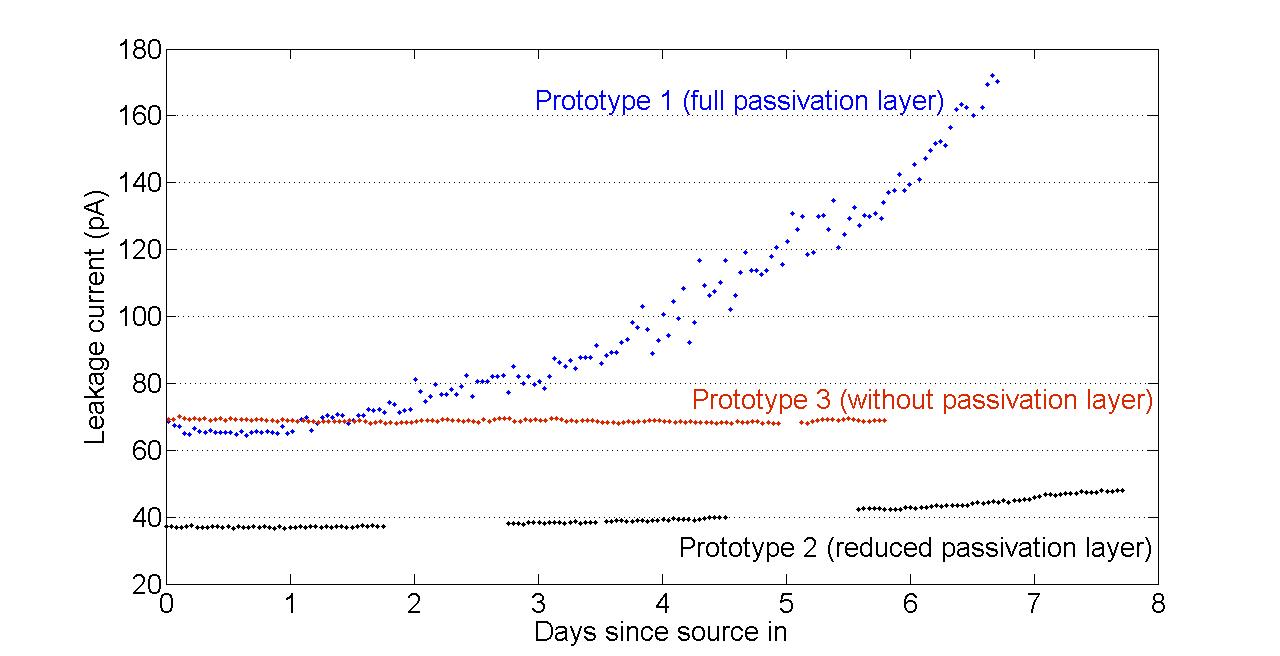}
	\caption{Leakage current in function of days of $\gamma$ irradiation for 3 detectors with different passivation layers. The LC increase is strongly reduced for Prototype 2 and no increase is observed with Prototype 3.}
	\label{fig:ircomp}
	\end{centering}
\end{figure}

To investigate the role of the passivation layer in the $\gamma$-radiation induced increase of LC, two other prototype detectors with different passivation layer geometries have been tested in LAr under $\gamma$ irradiation. Prototype 2 had a reduced passivation layer limited to the groove area and Prototype 3 had no passivation layer evaporated. Figure \ref{fig:ircomp} compares the $\gamma$-radiation induced LC of the three prototype detectors in LAr. The $\gamma$-radiation induced LC of the Prototype 2 is strongly suppressed compared to the first prototype (10 pA/week vs. 80 pA/week). The third prototype without passivation layer shows no increase of the LC even after one week of irradiation.

The most likely explanation to the $\gamma$-radiation induced LC is the collection and trapping of charge, produced by ionization of LAr, on the surface of the detector passivation layer. The electric field in the surrounding of the detector drifts the charge to the detector surface. Positive (negative) charges are collected on the inner (outer) part of the passivation layer, changing the conductivity of the surface, or in the bulk close to the surface. Electric field calculations show that more charges are collected on the detector surface with - HV than with + HV.

\begin{figure}[h]
		\begin{centering}
		\includegraphics[scale=0.2]{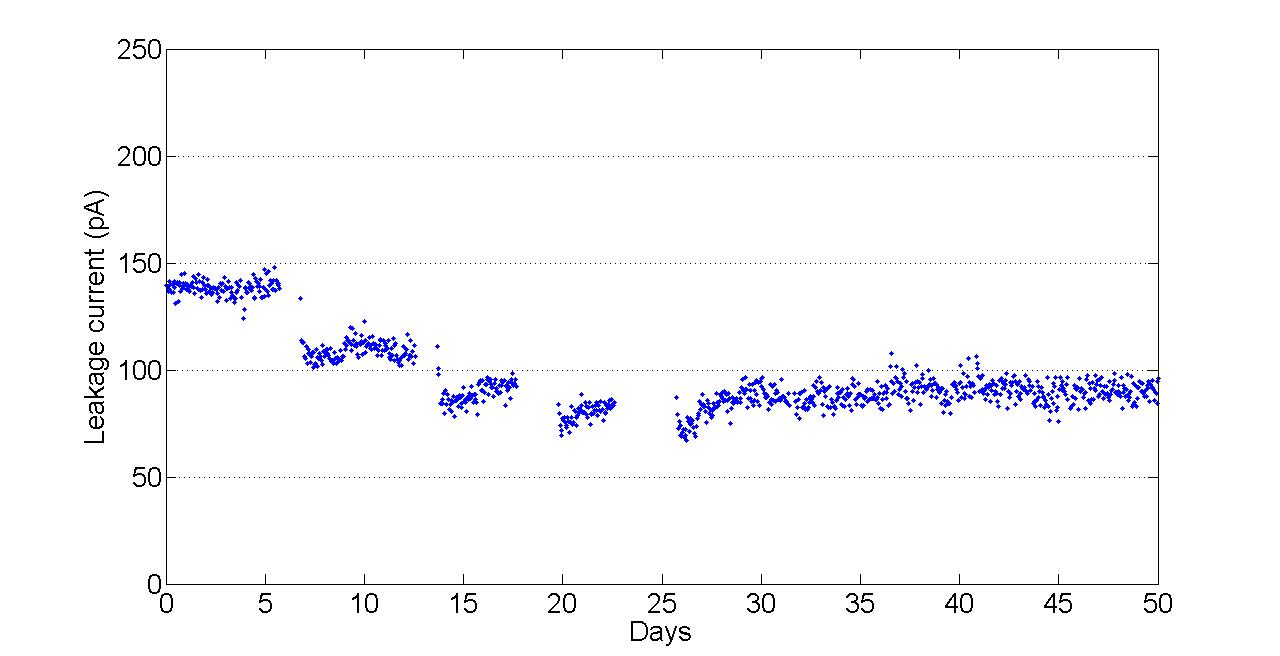}
	\caption{Gamma irradiations without applying HV result in a decrease of the LC. The HV is applied in between the irradiations to measure the LC.}
	\label{fig:nohv}
	\end{centering}
\end{figure}

The $\gamma$-radiation induced LC is a reversible process. First, after $\gamma$ irradiation of the HPGe diode in LAr without applying HV, a decrease of the LC is observed. Figure \ref{fig:nohv} shows the decrease of the LC after four consecutive $\gamma$ irradiations without applying HV.
The UV scintillation photons from LAr, breaking the bonds between ions and the passivation layer, are thought to be the curing agent. Second, after a warming/cooling cycle, the LC of the detector is restored to its original value.

\subsubsection{Long-term stability}
\begin{figure}[h]
		\begin{centering}
		\includegraphics[scale=0.2]{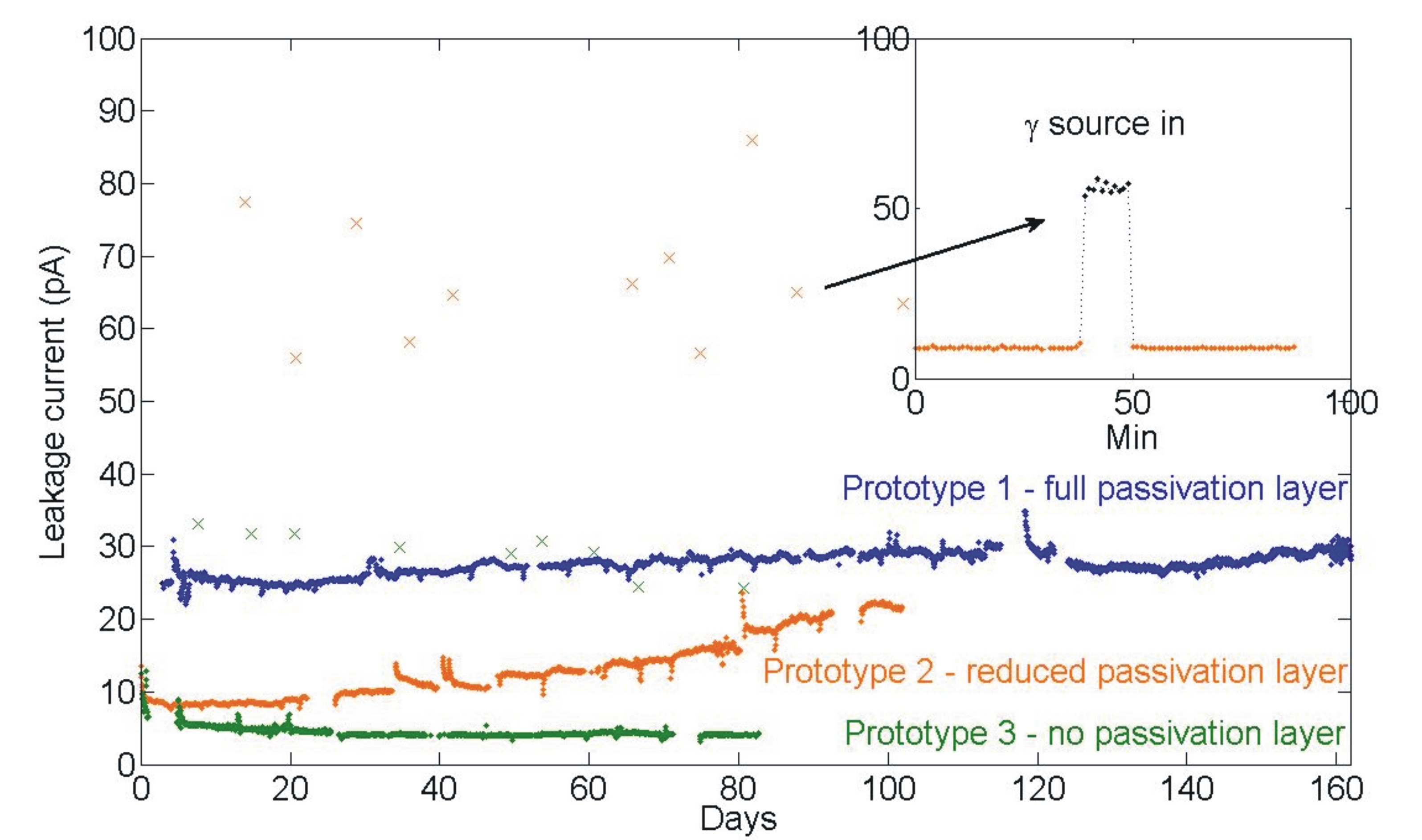}
	\caption{Long-term stability measurements in LAr with the three prototype detectors. Prototype 1 mounted with a PTFE/Cu/PTFE disk covering its passivation layer, Prototype 2 and Prototype 3 were operated for seven, four and three months respectively. Prototype 2 and 3 were exposed to a $^{60}$Co source once a week during 10 minutes.}
	\label{fig:long1}
	\end{centering}
\end{figure}

The success of GERDA depends strongly on the long-term stability of the detector parameters in LAr. To mitigate the $\gamma$-radiation induced LC, a PTFE/Cu/PTFE disk covering the passivation layer has been mounted on the Prototype 1 and the detector was continuously operated in LAr during seven months. Long-term stability measurements were also performed with Prototypes 2 and 3 for four and three months respectively. To mimic the energy calibration in GERDA, Prototypes 2 and 3 were exposed to $\gamma$-radiation once a week for 10-minute periods. Figure \ref{fig:long1} presents the long-term stability measurements in LAr for the three prototype detectors. The LC was continuously monitored with high accuracy amperemeters.
Prototype 1 and 3 show high stability but an increase of the LC is observed with Prototype 2. The small fluctuations of the LC are attributed to the LAr fillings once a week. For comparison, all detectors were biased during the measurements at the same HV of 4000 V, which is above their nominal bias voltage. 

\subsection{Phase-I detectors}
In 2006, the cryostats in which the enriched detectors were operated during HdM and IGEX experiments were opened and the dimensions of the diodes measured. The detectors were stored under vacuum while tests and measurements with prototype detectors were performed. In 2008, the detectors have been reprocessed at the detector manufacturer. As Prototype 3 shows the best performance in LAr, all Phase-I detectors have been reprocessed without the evaporation of a passivation layer. The detectors were returned to GDL and have been mounted in their low-mass holders. Presently, the characterization of the detectors in the LAr test stand of GDL is ongoing. Then, the detectors will be stored underground under vacuum until their operation in GERDA.

\section{Conclusion}
We have tested the operation and performance of bare HPGe detectors in LN$_2$ and in LAr over more than three years. The detector handling and mounting procedures have been defined and the Phase-I detector technology and the low-mass assembly have been tested successfully. The energy resolution of the first Phase-I prototype detector is 2.2 keV (FWHM) at 1.332 MeV with short signal and HV cables, identical to its performance in a standard vacuum cryostat, and 2.5 keV with $\sim$60 cm cables in the test bench in the Gerda underground Detector Laboratory (GDL) at LNGS. Studies with this prototype detector operated in LAr under varying irradiation conditions with a $\gamma$ source showed that long-term irradiation results in an increase of the LC. This effect is not observed when operating the same detector in LN$_2$. The LC increase is induced by charge collection on the passivation layer, which is however reversible. Gamma irradiation without biasing the detector, or thermal cycling the detector reestablishes the LC to its initial values. Stable LC values under $\gamma$ irradiation in LAr were observed with the prototype detector without passivation layer. All detector parameters are stable during the long-term measurements. The results concerning limited long-term stability of HPGe detectors operated in LN$_2$ reported in \cite{GTFfinal} are not confirmed.


%

\section*{Acknowledgment}

The authors would like to thank Jan Verplancke and Patrick Vermeulen, Canberra Semiconductor NV, Olen, for fruitfull discussions.

\ifCLASSOPTIONcaptionsoff
  \newpage
\fi

\end{document}